\documentclass[10pt,aps,prd,floatfix,notitlepage,showpacs]{revtex4-1}

\usepackage[utf8x]{inputenc}
\usepackage{amsmath,amssymb}

\begin{document}

\title{Variational approach to gravitational theories with two independent connections}
\author{Nicola Tamanini}
\affiliation{Department of Mathematics, University College London,
        Gower Street, London, WC1E 6BT, UK}
\email{n.tamanini.11@ucl.ac.uk}
\date{\today}

\begin{abstract}
A new variational approach for general relativity and modified theories of gravity is presented. In addition to the metric tensor, two independent affine connections enter the action as dynamical variables. In the matter action the dependence upon one of the connections is left completely unspecified. When the variation is applied to the Einstein-Hilbert action the Einstein field equations are recovered. However when applied to $f(R)$ and Scalar-Tensor theories, it yields gravitational field equations which differ from their equivalents obtained with a metric or Palatini variation and reduce to the former ones only when no connections appear in the matter action.
\end{abstract}


\maketitle


\section{Introduction}

The variational method of general relativity is not unique. The most popular variational approaches leading to the Einstein field equations when applied to the so-called Einstein-Hilbert action, are known under the names of metric and Palatini variational principles.

The first one, discovered by Hilbert at the very time Einstein was completing his theory, considers the metric tensor as the object containing all the gravitational degrees of freedom. It is the most common variational approach and is by now exposed in almost all (advanced) general relativity books (see e.g.~\cite{Wal84,MTW73}). The second one, although often attributed to Palatini because of \cite{Pal19}, has been developed by Einstein himself some years later (see \cite{FFR82} for a historical account). It consists of an independent variation of the action with respect to the metric and a torsionless connection (see again \cite{Wal84,MTW73}). Though it permits to relax the assumptions of the theory enlarging its geometrical setting (the connection is not postulated to be the Levi-Civita one), it also constrains the matter action to not depend upon the dynamical degrees of freedom represented by the connection, which is somehow theoretically unsatisfactory since it is well-known that the Fermionic matter action naturally contains the connection.

A third variational principle, known with the appelative of metric-affine (see \cite{Hehl:1994ue} for a review), considers both a non-symmetric connection and a fully general matter action from the beginning. However the equations of motion obtained varying independently with respect to the metric and the connection, do not reduce to the Einstein field equations unless some further assumptions are considered. Notably, for a matter action containing spinor fields the resulting theory differs from general relativity since a torsion-matter coupling arises.

Recently all these approaches have been applied to modified theories of gravity as well and produced a large amount of works, see \cite{DT10,Sotiriou:2008rp,Olmo:2011uz} for some reviews. It is well-known that different variations give rise to different physical theories when the gravitational action is not taken to be the Einstein-Hilbert one. In particular the simplest modifications one can imagine go under the name of $f(R)$ and Scalar-Tensor gravities, where the gravitational Lagrangian is modified by considering an arbitrary function of the Ricci scalar or a coupling between this and a scalar field, respectively. Remarkably, it has been proved that $f(R)$ gravity in both metric and Palatini approaches can be related to Brans-Dicke theories via conformal transformations \cite{Sotiriou:2008rp}.

Furthermore, more recently, new variational principles have been proposed and studied. A new bimetric variation has been analyzed in \cite{BeltranJimenez:2012sz} and metric and Palatini approaches have been unified as limits of a single variational method \cite{Amendola:2010bk,Koivisto:2011vq,Koivisto:2011tp} which can be applied to $f(R)$ modified gravity as well. On one side, these new works are motivated by the desire of finding new formulations of general relativity which could, for example, suggest useful hints for a consistent quantization of the theory. On the other hand, they can be employed trying to make sense out of the cosmological recent discoveries, i.e.~to built satisfactionary models of dark energy, dark matter and inflation.
No matter the reasons, it is historically well-known that new formulations of a given theory inevitably lead to further, physical or mathematical, insights.

In what follows we consider a new variational approach to gravitational theories where two completely independent connections are considered and no a priori assumptions are made on them. Together with the metric, they both enter as dynamical variables in the theory. The variation has to be taken independently with respect to the metric and both the connections. We apply this approach first to the Einstein-Hilbert action in Sec.~\ref{Sec:GR}, i.e.~to general relativity, then to $f(R)$ and Scalar-Tensor theories in Sec.~\ref{Sec:ModGrav}. We summarize the results and draw conclusions in Sec.~\ref{Sec:conc}.

\section{Biconnection General Relativity}
\label{Sec:GR}

Consider to generalize general relativity (GR) taking two independend connections $\Gamma_{\mu\nu}^\lambda$ 
and $\Omega_{\mu\nu}^\lambda$, both containing torsion and non-metricity.
The metric tensor $g_{\mu\nu}$ is still assumed to be a symmetric 2nd-rank tensor as in GR.
We will denote the Levi-Civita connection (which is completely determined by the metric) simply with
\begin{equation}
\left\{_{\mu\nu}^{\,\,\lambda}\right\} = \frac{1}{2} g^{\lambda\sigma} \left(\partial_\mu g_{\nu\sigma}+\partial_\nu g_{\mu\sigma} -\partial_\sigma g_{\mu\nu}\right) \,.
\end{equation}

We generalize the metric-affine gravitational action 
\begin{equation}
 S_{MA}(g,\Gamma) \equiv \frac{1}{2}\int d^4x \sqrt{-g}\,g^{\mu\nu} R_{\mu\nu}(\Gamma) =  \frac{1}{2}\int d^4x \sqrt{-g}\,R(\Gamma) \,,
 \label{002}
\end{equation}
where $R(\Gamma)$ is the Ricci scalar formed with the independent connection,
with the following action
\begin{eqnarray}
 S_{BC}(g,\Gamma,\Omega) &\equiv& \frac{1}{4}\int d^4x \sqrt{-g}\,g^{\mu\alpha}\delta_\nu^\beta \left[\partial_\mu\Gamma_{\alpha\nu}^\beta
    -\partial_\nu\Omega_{\alpha\mu}^\beta +\Gamma_{\lambda\mu}^\beta\Omega_{\alpha\nu}^\lambda
    -\Gamma_{\lambda\nu}^\beta\Omega_{\alpha\mu}^\lambda \right. \nonumber\\
    && \qquad\qquad \left. +\partial_\mu\Omega_{\alpha\nu}^\beta
    -\partial_\nu\Gamma_{\alpha\mu}^\beta +\Omega_{\lambda\mu}^\beta\Gamma_{\alpha\nu}^\lambda 
    -\Omega_{\lambda\nu}^\beta\Gamma_{\alpha\mu}^\lambda \right] \label{036} \\
 &\equiv& \frac{1}{4}\int d^4x\sqrt{-g}\,\, \Re(\Gamma,\Omega) \label{001} \,,
\end{eqnarray}
where we have defined
\begin{eqnarray}
 {\Re_{\mu\nu\alpha}}^\beta(\Gamma,\Omega) &\equiv& \partial_\mu\Gamma_{\alpha\nu}^\beta
    -\partial_\nu\Omega_{\alpha\mu}^\beta +\Gamma_{\lambda\mu}^\beta\Omega_{\alpha\nu}^\lambda
    -\Gamma_{\lambda\nu}^\beta\Omega_{\alpha\mu}^\lambda \nonumber\\
    && +\partial_\mu\Omega_{\alpha\nu}^\beta
    -\partial_\nu\Gamma_{\alpha\mu}^\beta +\Omega_{\lambda\mu}^\beta\Gamma_{\alpha\nu}^\lambda 
    -\Omega_{\lambda\nu}^\beta\Gamma_{\alpha\mu}^\lambda \,, \\
 \Re_{\mu\alpha}(\Gamma,\Omega) &\equiv& {\Re_{\mu\beta\alpha}}^\beta(\Gamma,\Omega) \,, \\
\Re(\Gamma,\Omega)&\equiv&g^{\mu\alpha}\Re_{\mu\alpha}(\Gamma,\Omega) \,,
\end{eqnarray}
in analogy with the Riemann tensor and its contractions.

We note that $ {\Re_{\mu\nu\alpha}}^\beta(\Gamma,\Omega)$ can be obtained generalizing the definition of the Riemann tensor, i.e.~considering a particular 
commutation of covariant derivatives mixing both $\Gamma$ and $\Omega$. For any vector $V^\mu$, we find
\begin{eqnarray}
 \frac{1}{2} \left(\stackrel{\Omega}{\nabla}_\mu\stackrel{\Gamma}{\nabla}_\nu
      -\stackrel{\Gamma}{\nabla}_\nu\stackrel{\Omega}{\nabla}_\mu
      +\stackrel{\Gamma}{\nabla}_\mu\stackrel{\Omega}{\nabla}_\nu
      -\stackrel{\Omega}{\nabla}_\nu\stackrel{\Gamma}{\nabla}_\mu \right)V^\beta
 = \left[ \frac{1}{2}{\Re_{\mu\nu\alpha}}^\beta(\Gamma,\Omega)
   +\delta_\alpha^\beta\left({S_{\mu\nu}}^\lambda(\Gamma) \stackrel{\Omega}{\nabla}_\lambda
                             +{S_{\mu\nu}}^\lambda(\Omega)\stackrel{\Gamma}{\nabla}_\lambda \right) \right]V^\alpha \,, &&
\label{022}
\end{eqnarray}
where
\begin{eqnarray}
 {S_{\mu\nu}}^\lambda(\Gamma) &\equiv& \Gamma_{[\mu\nu]}^\lambda \,=\, \frac{1}{2}\left( \Gamma_{\mu\nu}^\lambda-\Gamma_{\nu\mu}^\lambda \right) \,, \\
 {S_{\mu\nu}}^\lambda(\Omega) &\equiv& \Omega_{[\mu\nu]}^\lambda \,=\, \frac{1}{2}\left( \Omega_{\mu\nu}^\lambda-\Omega_{\nu\mu}^\lambda \right) \,,
\end{eqnarray}
are the torsion tensors of $\Gamma_{\mu\nu}^\lambda$ and $\Omega_{\mu\nu}^\lambda$, respectively.
Of course $\stackrel{\Gamma}{\nabla}$ is the covariant derivative with respect to $\Gamma_{\mu\nu}^\lambda$ and $\stackrel{\Omega}{\nabla}$ is the covariant derivative with respect to $\Omega_{\mu\nu}^\lambda$. Notice that taking $\Gamma=\Omega$ in (\ref{022}) gives back the usual metric-affine definition of the Riemann tensor (which generally includes a torsion term \cite{Hammond:2002rm}).

It is also immediate to notice that $S_{BC}$ is invariant under the exchange of the two connections, $\Gamma\rightleftarrows\Omega$. Moreover we have that $\Re(\Gamma,\Gamma)=R(\Gamma)$.
Thus, when we take $\Gamma=\Omega$ the action (\ref{001}) reduces to the standard metric-affine action (\ref{002}).
Another transformation leaving action (\ref{001}) unchanged is the so-called {\it projective transformation} \cite{Sandberg:1975db}, which in our case has to be performed on both the connections:
\begin{eqnarray}
 \Gamma_{\mu\nu}^\lambda &\rightarrow& \Gamma_{\mu\nu}^\lambda+\lambda_\mu\delta_\nu^\lambda \,, \label{011} \\
 \Omega_{\mu\nu}^\lambda &\rightarrow& \Omega_{\mu\nu}^\lambda+\lambda_\mu\delta_\nu^\lambda \,, \\
 g_{\mu\nu} &\rightarrow& g_{\mu\nu} \,, \label{012}
\end{eqnarray}
with $\lambda_\mu(x)$ denoting four spacetime functions. This can be easily realized rewriting action (\ref{001}) in the form
\begin{equation}
 S_{BC}=\frac{1}{4}\int d^4x\sqrt{-g}\left[R(\Gamma)+R(\Omega)-2g^{\mu\nu}{\mathcal{K}_{\nu[\mu}}^\sigma 
        {\mathcal{K}_{\sigma\beta]}}^\beta\right] \,.
\label{013}
\end{equation}
where we define the {\it difference tensor}
\begin{equation}
 {\mathcal{K}_{\mu\nu}}^\lambda \equiv \Gamma_{\mu\nu}^\lambda-\Omega_{\mu\nu}^\lambda \,.
\end{equation}
The (gauge) transformation (\ref{011})-(\ref{012}) implies that among all the 64 degrees of freedom of $\Gamma_{\mu\nu}^\lambda$ (or equivalently $\Omega_{\mu\nu}^\lambda$) only 60 can be fully determined. We will always have 4 degrees of freedom that cannot be fixed by the equations of motion. These are to be fixed introducing a Lagrange multiplier as we are going to understand soon.

At this point we add to $S_{BC}$ the matter action $S_{M}$. In the matter action we have to make a choice upon the connection used in the covariant derivative of the matter fields. We assume that the matter action depends only by one connection, or, in other words, the GR minimal coupling has to be performed using the same connection for all matter fields, which is a reasonably physical requirement. Because of the symmetry $\Gamma\rightleftarrows\Omega$ in $S_{BC}$, it does not matter which connection among $\Gamma$ or $\Omega$ will appear in $S_{M}$. We choose $\Gamma$.

In the action
\begin{equation}
S\equiv S_{BC}(g,\Gamma,\Omega)+S_{M}(g,\Gamma,\Psi)\,,
\label{026}
\end{equation}
where $\Psi$ denotes all matter fields collectively, the projective transformation (\ref{011})-(\ref{012}) is now broken because of the matter term. This means that we must fix four degrees of freedom in the biconnection action, otherwise inconsistencies would arise in the field equations. This can be achieved introducing a Lagrange multiplier $B^\mu$ constraining part of $\Gamma^\lambda{}_{\mu\nu}$. Following \cite{Sandberg:1975db} we can add to action (\ref{026}) the term
\begin{equation}
S_{LM} = \int d^4x \sqrt{-g} B^\mu S_{\lambda\mu}{}^\lambda(\Gamma) \,,
\end{equation}
which sets to zero the four degrees of freedom of $\Gamma$ represented by $S_{\lambda\mu}{}^\lambda(\Gamma)$.

We are now ready to vary the total action
\begin{equation}
S\equiv S_{BC}(g,\Gamma,\Omega)+S_{M}(g,\Gamma,\Psi)+S_{LM}(g,\Gamma,B)\,.
\label{023}
\end{equation}
The variation with respect to the Lagrange multiplier $B^\mu$ yields of course
\begin{equation}
S_{\lambda\mu}{}^\lambda(\Gamma)=0 \,.
\label{024}
\end{equation}
In order to vary with respect to $\Omega$ and $\Gamma$ we rewrite the biconnection action $S_{BC}$ in a more practical form as
\begin{equation}
 S_{BC}=\frac{1}{2}\int d^4x \sqrt{-g}\,g^{\mu\nu} \left[R_{\mu\nu}(\Gamma)
        -\stackrel{\Gamma}{\nabla}_{[\mu}{\mathcal{K}_{\nu\beta]}}^\beta
        -{S_{\mu\beta}}^\lambda(\Gamma)\,{\mathcal{K}_{\nu\lambda}}^\beta \right] \,,
\label{008}
\end{equation}
where we have taken advantage of the following useful relation
\begin{equation}
\frac{1}{2}\Re_{\mu\nu}=R_{\mu\nu}(\Gamma)
        -\stackrel{\Gamma}{\nabla}_{[\mu}{\mathcal{K}_{\nu\beta]}}^\beta
        -{S_{\mu\beta}}^\lambda(\Gamma)\,{\mathcal{K}_{\nu\lambda}}^\beta \,.
\label{044}
\end{equation}

Let vary (\ref{008}) with respect to $\Omega_{\mu\nu}^\lambda$. We obtain the equations (using tensor densities covariant differentiation rules)
\begin{eqnarray}
-\stackrel{\Gamma}{\nabla}_\beta\Big(\sqrt{-g}\,g^{\mu\nu}\Big)
+\stackrel{\Gamma}{\nabla}_\lambda\left(\sqrt{-g}\,g^{\lambda\nu}\right) \delta_\beta^{\mu}+ 
2\,\sqrt{-g}\left[ g^{\lambda\nu}{S_{\lambda\beta}}^{\mu}(\Gamma) +g^{\mu\nu}{S_{\beta\sigma}}^\sigma(\Gamma) -g^{\lambda\nu}\delta_\beta^\mu {S_{\lambda\sigma}}^\sigma(\Gamma) \right] =0 \,,
\label{009}
\end{eqnarray}
which are exactly the same we would obtain varying the Einstein-Hilbert action with respect to the connection within the Metric-Affine approach while assuming the matter action does not depend on $\Gamma$.
They can be rewritten as \cite{Utrecth2006}
\begin{equation}
{P^{\mu\nu}}_{\alpha\beta}\left(\partial_\lambda\, g^{\alpha\beta} +\tilde\Gamma_{\sigma\lambda}^\alpha g^{\sigma\beta} +\tilde\Gamma_{\lambda\sigma}^\beta g^{\alpha\sigma} \right) =0\,,
\label{010}
\end{equation}
where we have introduced the new connection
\begin{equation}
\tilde\Gamma_{\mu\nu}^\lambda\equiv \Gamma_{\mu\nu}^\lambda-\frac{2}{3}\delta_\nu^\lambda {S_{\mu\sigma}}^\sigma \,,
\label{035}
\end{equation}
and we have defined the tensor
\begin{equation}
{P^{\mu\nu}}_{\alpha\beta} \equiv \delta_\alpha^\mu\delta_\beta^\nu -\frac{1}{2}g_{\alpha\beta}g^{\mu\nu} \,.
\label{034}
\end{equation}
Thanks to (\ref{024}), the connection (\ref{035}) immediately reduces to $\Gamma$, namely
\begin{equation}
\tilde\Gamma_{\mu\nu}^\lambda\equiv \Gamma_{\mu\nu}^\lambda \,,
\end{equation}
and (\ref{010}) implies
\begin{equation}
\partial_\lambda\, g_{\alpha\beta} -\Gamma_{\sigma\lambda}^\sigma g_{\sigma\beta} -\Gamma_{\lambda\beta}^\sigma g_{\alpha\sigma} =0 \,.
\end{equation}
This can be algebraically solved to give \cite{Sandberg:1975db,Utrecth2006,Scrodinger}
\begin{equation}
\Gamma_{\mu\nu}^\lambda = \left\{_{\mu\nu}^{\,\,\lambda}\right\} \,,
\label{025}
\end{equation}
implying that $\Gamma$ in nothing but the Levi-Civita connection.

Before varying $S$ with respect to $\Gamma_{\mu\nu}^\lambda$, we define the {\it hypermomentum} $\Delta$ as the variation of the matter Lagrangian density $\mathcal L_M$ with respect to $\Gamma$ \cite{Hehl:1978}, i.e.
\begin{equation}
{\Delta_\mu}^{\lambda\beta}\equiv -\frac{1}{\sqrt{-g}}\frac{\delta\mathcal L_M}{\delta\Gamma_{\lambda\beta}^\mu} \,.
\end{equation}
In this manner performing the variation of $S$ in $\Gamma$, and after having taking into account field equations (\ref{009}), we arrive at
\begin{equation}
g^{\alpha\beta} {\mathcal{K}_{\beta\alpha}}^\nu \delta_\sigma^\mu + g^{\mu\nu}{\mathcal{K}_{\sigma\beta}}^\beta -g^{\lambda\nu} {\mathcal{K}_{\sigma\lambda}}^{\mu} -g^{\mu\lambda} {\mathcal{K}_{\lambda\sigma}}^{\nu} + 4B^{[\mu}\delta^{\nu]}_\sigma = 4\, {\Delta_\sigma}^{\nu\mu} \,.
\label{027}
\end{equation}
Contracting with respect to the indices $\nu$ and $\sigma$, the Lagrange multiplier $B^\mu$ can be related to the Hypermomentum as
\begin{equation}
B^\mu = \frac{2}{3} \Delta_\nu{}^{\nu\mu} \,.
\end{equation}
Using this back in (\ref{027}) leads to
\begin{equation}
g^{\alpha\beta} {\mathcal{K}_{\beta\alpha}}^\nu \delta_\sigma^\mu + g^{\mu\nu}{\mathcal{K}_{\sigma\beta}}^\beta -g^{\lambda\nu} {\mathcal{K}_{\sigma\lambda}}^{\mu} -g^{\mu\lambda} {\mathcal{K}_{\lambda\sigma}}^{\nu} = 4\, {H_\sigma}^{\nu\mu} \,,
\label{006}
\end{equation}
where we have defined
\begin{equation}
{H_\sigma}^{\nu\mu} \equiv {\Delta_\sigma}^{\nu\mu}- \frac{2}{3}\Delta_\lambda{}^{\lambda[\mu} \delta^{\nu]}_\sigma \,.
\end{equation}
Equation (\ref{006}) will turn out useful to simplify the other field equations later on. Note that (\ref{006}) is a system of algebraic equations in $\mathcal K_{\mu\nu}{}^\lambda$, meaning that whenever $\Delta_\lambda{}^{\mu\nu}=0$ we have the solution $\Gamma_{\mu\nu}{}^\lambda=\Omega_{\mu\nu}{}^\lambda$ and the consequent reduction of $S_{BC}$ to the Palatini action.

Finally, the variation of action (\ref{023}) with respect to the metric $g^{\mu\nu}$, after taking into account field equation (\ref{024}), yields
\begin{equation}
{G}_{(\alpha\beta)}(\Gamma)+\frac{1}{2}{P^{\mu\nu}}_{(\alpha\beta)} \left(\stackrel{\Gamma}{\nabla}_\lambda{\mathcal{K}_{\nu\mu}}^\lambda -\stackrel{\Gamma}{\nabla}_\mu{\mathcal{K}_{\nu\lambda}}^\lambda -2{S_{\mu\sigma}}^\lambda(\Gamma){\mathcal{K}_{\nu\lambda}}^\sigma\right) =\; \bar{T}_{\alpha\beta}(\Gamma) \,,
\label{005}
\end{equation}
where
\begin{equation}
{G}_{\mu\nu}(\Gamma) \equiv R_{\mu\nu}(\Gamma)-\frac{1}{2}g_{\mu\nu}R(\Gamma) \,,
\end{equation}
is the Einstein tensor in terms of $\Gamma$ and $P^{\mu\nu}{}_{\alpha\beta}$ is given in (\ref{034}).
The stress-energy tensor appearing in the RHS of (\ref{005}) is defined by
\begin{equation}
\bar{T}_{\mu\nu}(\Gamma) \equiv -\frac{2}{\sqrt{-g}} \frac{\delta\mathcal{L}_M}{\delta g^{\mu\nu}} \,,
\end{equation}
and differs from the usual stress-energy tensor $T_{\mu\nu}$ inasmuch as it lacks of the variation of $\left\{_{\mu\nu}^{\,\,\lambda}\right\}$ in $g^{\mu\nu}$. Note that in principle $\bar T_{\mu\nu}$ still depends on $\Gamma$ if this appears not only linearly in the matter action. A discussion about this issue will be presented in Sec.~\ref{Sec:ModGrav} where it will become of central importance.

Equations (\ref{005}) can be further simplified recalling condition (\ref{025}) which allows to rewrite it as
\begin{equation}
G_{\alpha\beta}+\frac{1}{2}{P^{\mu\nu}}_{(\alpha\beta)} \left({\nabla}_\lambda{\mathcal{K}_{\nu\mu}}^\lambda -{\nabla}_\mu{\mathcal{K}_{\nu\lambda}}^\lambda\right) =\; \bar{T}_{\alpha\beta} \,,
\end{equation}
where now $G_{\mu\nu}$ is the usual Einstein tensor formed with the Levi-Civita connection (which is also symmetric by definition) and $\nabla$ represents the covariant derivative taken with respect to the Levi-Civita connection. Of course also $\bar T_{\mu\nu}$ depends now on the Levi-Civita connection, but it still differs from $T_{\mu\nu}$ because of the missing variation of the connection.

At this point we define \cite{Hehl:1978}
\begin{equation}
\sigma^{\beta\mu\nu} \equiv \Delta^{\beta(\mu\nu)} =H^{\beta(\mu\nu)} \,.
\end{equation}
Then, using (\ref{006}), after some algebra we find
\begin{equation}
{P^{\mu\nu}}_{(\alpha\beta)} \left({\mathcal{K}_{\nu\mu}}^\lambda -{\mathcal{K}_{\nu\sigma}}^\sigma\delta_\mu^\lambda\right) = -2\left({\sigma_{\beta\alpha}}^\lambda +\sigma_{\alpha\;\;\beta}^{\;\;\lambda} -{\sigma^\lambda}_{\beta\alpha}\right) \,,
\end{equation}
which inserted into (\ref{005}) gives
\begin{eqnarray}
G^{\alpha\beta} &=& \bar{T}^{\alpha\beta} +\nabla_\lambda \left(\sigma^{\beta\alpha\lambda}+\sigma^{\alpha\lambda\beta} -\sigma^{\lambda\beta\alpha}\right) \,, \nonumber\\
&=& \bar{T}^{\alpha\beta} +\nabla_\lambda \left(\Delta^{\beta(\alpha\lambda)}+\Delta^{\alpha(\lambda\beta)} -\Delta^{\lambda(\beta\alpha)}\right) \label{007} \,.
\end{eqnarray}
The divergence in the RHS of these equations represents exactly the missing variation of the connection with respect to the metric in the stress-energy tensor $\bar T^{\mu\nu}$ as it is explicitly shown in the Appendix (see also \cite{Hehl:1978}). In other words (\ref{007}) simply reduce to
\begin{equation}
G_{\mu\nu}=T_{\mu\nu} \,,
\end{equation}
which are nothing but the usual {\it Einstein field equations}.

In conclusion, the biconnection formulation of GR presented here, is physically equivalent to the usual metric formulation. It differs from the Palatini formulation inasmuch as it allows for a generalized dependence upon the connection in the matter action and does not assume any a priori conditions on this connection. In this respect, it is more similar to the metric-affine variation except that at the end of the day it yields to the usual Einstein field equations. We can think of biconnection GR as a different formulation of GR rather than a new physical theory. The only problem it permits to solve is the general dependence in the matter action of the connection. As we are now going to see, for different (modified) gravitational theories the biconnection approach yields to some physical deviations from the usual Palatini variation.

\section{Biconnection Modified Gravitational Theories}
\label{Sec:ModGrav}

Let now generalize action (\ref{001}) with the following action
\begin{equation}
S_{MG} \equiv \frac{1}{4} \int d^4x \sqrt{-g} \, f(\phi,\Re) \,,
\label{029}
\end{equation}
where $f$ is an arbitrary function of both a scalar field $\phi$ and the biconnection curvature scalar $\Re(\Gamma,\Omega)$. This action reduces to the (biconnection formulation of) $f(R)$ theories whenever $f$ does not depend on $\phi$ and to Scalar-Tensor thoeries when $f=W(\phi)\Re$ for some function $W$ of $\phi$, i.e.~if $f$ happens to be linear in $\Re$. Several studies have already considered action (\ref{029}) in the contest of both metric and Palatini variations (see e.g.~\cite{Tamanini:2010uq, Koivisto:2005yc, Hwang:2001qk, Capozziello:2007ec}). The theory contains both $f(R)$ theories and Scalar-Tensor theories as subcases, and the general features and results which follows will automatically hold in these two classes of theories too.

Again we want to vary the total action
\begin{equation}
S\equiv S_{MG}(g,\Gamma,\Omega,\phi)+S_{M}(g,\Gamma,\phi,\Psi)+S_{LM}(g,\Gamma,B)\,.
\label{015}
\end{equation}
independently with respect to $g_{\mu\nu}$, $B^\mu$, $\Gamma_{\mu\nu}^\lambda$ and $\Omega_{\mu\nu}^\lambda$ (the variation with respect to the scalar field $\phi$ is irrelevant in what follows and will thus not be explicitly considered). Note that we allow the scalar field $\phi$ to appear in the matter action where it is free to form a kinetic term, a potential term and even to couple with other matter fields.
Clearly the variation with respect to $B^\mu$ gives again (\ref{024}), i.e.
\begin{equation}
S_{\lambda\mu}{}^\lambda(\Gamma)=0 \,.
\label{028}
\end{equation}

Before proceeding we define the derivation of $f$ with respect to $\Re$ as
\begin{equation}
F(\phi,\Re) \equiv \frac{\partial f}{\partial \Re} \,.
\end{equation}
Note that in the case of Scalar-Tensor theories we have $F=W(\phi)$.
The variation in $\Omega$ yields then to the following equations of motion
\begin{eqnarray}
{S_{\lambda\sigma}}^\sigma(\Gamma)g^{\lambda\nu}\delta^\mu_\beta -{S_{\beta\sigma}}^\sigma(\Gamma)g^{\mu\nu} -{S_{\lambda\beta}}^\mu(\Gamma)g^{\lambda\nu}+  \frac{1}{2F\sqrt{-g}}\stackrel{\Gamma}{\nabla}_\beta \left(\sqrt{-g}Fg^{\mu\nu}\right) -\frac{1}{2F\sqrt{-g}}\stackrel{\Gamma}{\nabla}_\lambda \left(\sqrt{-g}Fg^{\lambda\nu}\right)\delta^\mu_\beta =0 \,,
\label{014}
\end{eqnarray}
which, in terms of the connection (\ref{035})
implies \cite{Olmo:2011uz,Sotiriou:2009xt}
\begin{equation}
\partial_\lambda\, g^{\alpha\beta} +\tilde\Gamma_{\sigma\lambda}^\alpha g^{\sigma\beta} +\tilde\Gamma_{\lambda\sigma}^\beta g^{\alpha\sigma} +\frac{\partial_\lambda F}{F}g^{\alpha\beta} =0\,.
\label{031}
\end{equation}
This, taking into account (\ref{028}), can be algebraically solved to obtain
\begin{equation}
\Gamma_{\mu\nu}^\lambda =\tilde\Gamma_{\mu\nu}^\lambda = \left\{_{\mu\nu}^{\,\,\lambda}\right\} +\frac{1}{2F}\left(\partial_\mu F\delta^\lambda_\nu +\partial_\nu F\delta^\lambda_\mu -g^{\lambda\sigma}g_{\mu\nu}\partial_\sigma F\right) \,.
\label{019}
\end{equation}
We then managed to explicitly find an expression for $\Gamma$ from the field equations which is given by (\ref{019}). Note that this expression is the same appearing in Palatini modified gravity \cite{Sotiriou:2008rp,Olmo:2011uz,Tamanini:2010uq,Koivisto:2005yc}.

Let now perform the variation in $\Gamma$. After having considered (\ref{014}) we arrive at the field equations
\begin{equation}
g^{\alpha\beta} {\mathcal{K}_{\beta\alpha}}^\nu \delta_\sigma^\mu + g^{\mu\nu}{\mathcal{K}_{\sigma\beta}}^\beta -g^{\lambda\nu} {\mathcal{K}_{\sigma\lambda}}^{\mu} -g^{\mu\lambda} {\mathcal{K}_{\lambda\sigma}}^{\nu} = \frac{4}{F}\left( {\Delta_\sigma}^{\nu\mu}- B^{[\mu}\delta^{\nu]}_\sigma \right)\,,
\label{030}
\end{equation}
which generalize equations (\ref{006}). Again a contraction on $\nu$ and $\sigma$ gives
\begin{equation}
B^\mu = \frac{2}{3} \Delta_\nu{}^{\nu\mu} \,,
\end{equation}
which inserted back in (\ref{030}) yields
\begin{equation}
g^{\alpha\beta} {\mathcal{K}_{\beta\alpha}}^\nu \delta_\sigma^\mu + g^{\mu\nu}{\mathcal{K}_{\sigma\beta}}^\beta -g^{\lambda\nu} {\mathcal{K}_{\sigma\lambda}}^{\mu} -g^{\mu\lambda} {\mathcal{K}_{\lambda\sigma}}^{\nu} = \frac{4}{F} {H_\sigma}^{\nu\mu}\,.
\label{020}
\end{equation}
Again, we will use (\ref{020}) to simplify the other field equations. Moreover we still recover Palatini when $\Delta_\lambda{}^{\mu\nu}=0$ since from (\ref{020}) we obtain $\Gamma = \Omega$ as in GR, meaning that in vacuum the two variations lead to the same physics. This issue will be further analyzed later on in this section.

Finally, varying action (\ref{015}) in the metric $g^{\mu\nu}$ and taking into account condition (\ref{028}), gives
\begin{equation}
F\,\Re_{(\mu\nu)}-\frac{1}{2}g_{\mu\nu}f=\bar{T}_{\mu\nu}(\Gamma) \,,
\label{016}
\end{equation}
which can be rewritten as
\begin{equation}
F{P^{\alpha\beta}}_{(\mu\nu)}\Re_{\alpha\beta} +\frac{1}{2}Fg_{\mu\nu}\Re -\frac{1}{2}g_{\mu\nu}f=\bar{T}_{\mu\nu}(\Gamma) \,.
\end{equation}
Expanding $\Re_{\mu\nu}$ with the help of (\ref{044}) leads to
\begin{eqnarray}
F{P^{\alpha\beta}}_{(\mu\nu)} R_{\alpha\beta}(\Gamma) 
+\frac{1}{2}Fg_{\mu\nu}\Re -\frac{1}{2}g_{\mu\nu}f +  F{P^{\alpha\beta}}_{(\mu\nu)} \left[\frac{1}{2}\stackrel{\Gamma}{\nabla}_\sigma {\mathcal{K}_{\beta\alpha}}^\sigma -\frac{1}{2}\stackrel{\Gamma}{\nabla}_\alpha {\mathcal{K}_{\beta\sigma}}^\sigma +{S_{\alpha\sigma}}^\lambda(\Gamma){\mathcal{K}_{\beta\lambda}}^\sigma\right] =\bar{T}_{\mu\nu}(\Gamma) \,.
\end{eqnarray}
Thanks to (\ref{019}) this becomes
\begin{align}
F\,\tilde G_{\mu\nu}+ F{P^{\alpha\beta}}_{(\mu\nu)} \left[\frac{1}{2}\tilde{\nabla}_\sigma {\mathcal{K}_{\beta\alpha}}^\sigma -\frac{1}{2}\tilde{\nabla}_\alpha {\mathcal{K}_{\beta\sigma}}^\sigma\right]  =\bar T_{\mu\nu}(\tilde\Gamma) -\frac{1}{2}F\,g_{\mu\nu}\Re +\frac{1}{2}g_{\mu\nu}f \,, \label{037}
\end{align}
where $\tilde G_{\mu\nu}$ and $\tilde\nabla$ are the Einstein tensor and the covariant derivative with respect to $\tilde\Gamma$.  
Note that ${S_{\mu\nu}}^\lambda(\tilde\Gamma)=0$, as can be easily seen from (\ref{019}).

A discussion has to be made about $\bar T_{\mu\nu}(\tilde\Gamma)$ at this point. This is still the variation of the matter action with respect to $g^{\mu\nu}$, but now depends on the new connection $\tilde\Gamma$. In general, since $\tilde\Gamma$ is given by the Levi-Civita connection plus a contribution depending on $F$ (see (\ref{019})), we can separate $\bar T_{\mu\nu}(\tilde\Gamma)$ in $\bar T_{\mu\nu}$, depending only on the metric being the connections replaced by the Levi-Civita ones, and $T^{(F)}_{\mu\nu}$, which depends on (derivatives of) $F$ and the metric. We do not know the explicit expression of $T^{(F)}_{\mu\nu}$ until the matter action is specified, however whenever $S_M$ depends only linearly on $\Gamma$ we have $T^{(F)}_{\mu\nu}=0$ since $\bar T_{\mu\nu}(\Gamma)$ becomes $\Gamma$ independent. This is a physically reasonable constraint since derivatives and quadratic terms in $\Gamma$ are usually allowed to appear only in the gravitational action and not in the matter one. In what follows we will assume $T^{(F)}_{\mu\nu}=0$, which means
\begin{equation}
\bar T_{\mu\nu}(\Gamma) = \bar T_{\mu\nu}(\tilde\Gamma)= \bar T_{\mu\nu} \,.
\label{042}
\end{equation}
Note that in biconnection GR we do not have to impose the constraint (\ref{042}) since it happens to be automatically satisfied thanks to (\ref{025}).
Moreover a peculiar feature distinguishing the biconnection GR action (\ref{036}) from all the other gravitational actions (including the Einstein-Hilbert one), is that it is linearly dependent through $\Gamma$ and its first derivatives. Thus, curiously enough, while in action (\ref{036}) $\Gamma$ appears only linearly, in biconnection GR we can allow the (matter) connection to form quadratic and higher order terms inside the matter action.

Let now look back at (\ref{016}), we notice that a contraction with $g^{\mu\nu}$ provides
\begin{equation}
F\,\Re-2f=\bar{T}(\Gamma) \,,
\label{038}
\end{equation}
where $\bar{T}(\Gamma)\,\equiv g^{\mu\nu}\bar{T}_{\mu\nu}(\Gamma)$. According to the argument above we can split (\ref{038}) as
\begin{equation}
F\,\Re-2f=\bar{T}+T^{(F)} \,,
\label{039}
\end{equation}
which considering constraint (\ref{042}) reduces to
\begin{equation}
F\,\Re-2f=\bar{T} \,.
\label{043}
\end{equation}
Note that $\Gamma$ has now disappeared from equation (\ref{043}).
Once given a particular form for $f$, equation (\ref{043}) gives $\Re$ in terms of $\bar{T}$ and $\phi$. This means that $\Re$, wherever appears in our field equations, can be substituted in favour of a source term and thus will not carry any spacetime dynamics (again this is in analogy with what happens in Palatini $f(R)$ gravity \cite{Sotiriou:2008rp}).
In general if we let $T^{(F)}_{\mu\nu}$ not to vanish, (\ref{039}) would become a differential equation for $\Re$, since $T^{(F)}$ would contain spacetime derivatives of $F$ and consequently of $\Re$. Then it would be hard to recover well-defined gravitational field equations, since the spacetime dynamics would become much more complicated.

Let now turn back to the derivation of the field equations.
Using (\ref{020}), equation (\ref{037}) can be rewritten as
\begin{multline}
\tilde G_{\mu\nu} -\frac{\partial_\delta F}{4F}P^{\alpha\beta}{}_{(\mu\nu)} \left(\delta_\beta^\delta\delta_\lambda^\sigma- g^{\delta\sigma}g_{\beta\lambda}\right) \left({\mathcal{K}_{\sigma\alpha}}^\lambda -{\mathcal{K}_{\sigma\rho}}^\rho\delta_\alpha^\lambda +{\mathcal{K}_{\alpha\sigma}}^\lambda -{\mathcal{K}_{\alpha\rho}}^\rho\delta_\sigma^\lambda \right) \\ =\frac{1}{F}\bar{T}_{\mu\nu} +\frac{1}{F}\nabla_\lambda \left(\Delta^{\beta(\alpha\lambda)}+\Delta^{\alpha(\lambda\beta)} -\Delta^{\lambda(\beta\alpha)}\right) -\frac{1}{2}g_{\mu\nu} \left(\Re-\frac{f}{F}\right) \\
=\frac{1}{F}\,{T}_{\mu\nu} -\frac{1}{2}g_{\mu\nu} \left(\Re-\frac{f}{F}\right) \,,
\end{multline}
where $T_{\mu\nu}$ is the usual energy-momentum of GR (see the Appendix).
In order to further simplify these equations we note that, using again (\ref{020}), we have (after a bit of algebra)
\begin{eqnarray}
P^{\alpha\beta}{}_{(\mu\nu)} \left(\delta_\beta^\delta\delta_\lambda^\sigma-g^{\delta\sigma}g_{\beta\lambda}\right) \left({\mathcal{K}_{\sigma\alpha}}^\lambda -{\mathcal{K}_{\sigma\rho}}^\rho\delta_\alpha^\lambda +{\mathcal{K}_{\alpha\sigma}}^\lambda -{\mathcal{K}_{\alpha\rho}}^\rho\delta_\sigma^\lambda \right) 
=-\frac{4}{F}\left({\sigma^\delta}_{\mu\nu} +\delta^\delta_{(\mu} \,{\sigma_{\nu)\lambda}}^\lambda \right) \,,
\end{eqnarray}
where $\sigma_{\lambda\mu\nu}\equiv\Delta_{\lambda(\mu\nu)}$. In this manner we get
\begin{eqnarray}
\tilde G_{\mu\nu}=\frac{1}{F}\,T_{\mu\nu} -\frac{1}{2}g_{\mu\nu}\left(\Re-\frac{f}{F}\right) -\frac{\partial_\lambda F}{F^2}\left({\sigma^\lambda}_{\mu\nu} -\delta^\lambda_{(\mu}{\sigma_{\nu)\alpha}}^\alpha\right) \,,
\end{eqnarray}
or, expanding $\tilde G_{\mu\nu}$,
\begin{multline}
G_{\mu\nu}=\frac{1}{F}\,T_{\mu\nu} -\frac{1}{2}g_{\mu\nu}\left(\Re-\frac{f}{F}\right) -\frac{\partial_\lambda F}{F^2}\left({\sigma^\lambda}_{\mu\nu} -\delta^\lambda_{(\mu}{\sigma_{\nu)\alpha}}^\alpha\right) \\
+\frac{1}{F}\left(\nabla_\mu\nabla_\nu-g_{\mu\nu}\Box\right)F -\frac{3}{2F^2}\left[\left(\nabla_\mu F\right)\left(\nabla_\nu F\right)-\frac{1}{2}g_{\mu\nu}\left(\nabla F\right)^2\right] \,.
\label{021}
\end{multline}
These are the gravitational field equations of biconnection modified gravity. They result similar to the gravitational field equations of Palatini $f(\phi,R)$ gravity and differ from them only through the terms containing $\sigma_{\lambda\mu\nu}$ and in the fact that $\Re$ becomes a function of $\bar{T}$ rather than $T$.
Note that if $f=\Re$ equations (\ref{021}) reduce to the usual Einstein field equations being thus consistent with the result found in Sec.~\ref{Sec:GR}.

In the case of $f(\Re)$ gravity we simply have $F=f'$ and the scalar field drops out from the field equations.
More interestingly, when we reduce the analysis to Scalar-Tensor theories, where $F=W(\phi)$, every contribution of $\bar{T}$ disappear from the field equations (\ref{021}) since $F$ will depend only on $\phi$ and
\begin{equation}
\left(\Re-\frac{f}{F}\right) = 0 \,.
\end{equation}
This means that for Scalar-Tensor theories there are no physical differences between biconnection and Palatini variations whenever $\sigma_{\lambda\mu\nu}=0$. This last condition is satisfied by all macroscopically known matter fields and even by fields giving rise to spin current, such as the Dirac field, which present only nonvanishing antisymmetric hypermomentum, $\tau_{\lambda\mu\nu}\equiv\Delta_{\lambda[\mu\nu]}$ \cite{Hehl:1994ue}.

As an example of a matter Lagrangian presenting non-vanishing hypermomentum we consider the dilaton scalar field $\psi$ of \cite{Burton:1997pe}, whose matter Lagrangian can be written as
\begin{align}
\sqrt{-g}\,\mathcal L_M = A(\psi)\, \partial_\mu\psi \partial^\mu\psi +B(\psi)\, g_{\mu\nu}\partial^\sigma\psi \stackrel{\Gamma}{\nabla}_\sigma g^{\mu\nu}  +C(\psi)\, \partial_\mu\psi \stackrel{\Gamma}{\nabla}_\nu g^{\mu\nu} +D(\psi) \stackrel{\Gamma}{\nabla}_\mu \partial^\mu\psi \,,
\end{align}
where $A$, $B$, $C$ and $D$ are arbitrary functions of $\psi$. Note that the connection $\Gamma$ enters the matter Lagrangian only linearly meaning that this is a suitable Lagrangian for our modified gravity analysis. The symmetric hypermomentum is given by
\begin{align}
\sigma_{\lambda\mu\nu} = (C-D)\left( g_{\mu\nu}\partial_\lambda\psi -4 g_{\lambda(\mu}\partial_{\nu)} \right) \,,
\end{align}
and vanishes only if $C=D$. Hence in this particular gravitational theory the field equations (\ref{021}) will in general differ from the ones obtained with a Palatini variation.

Exactly as it happens in the Palatini formulation, the biconnection approach reduces to the metric (and Palatini) approach only in the case of GR. For all the other gravitational theories it gives rise to different equations of motions in the presence of matter and reduce to the Palatini approach in vacuum.
In fact if the matter action does not depend on $\Gamma$, field equations (\ref{021}) become exactly the field equations of Palatini modified gravity since then $\sigma_{\lambda\mu\nu}=0$ and $\bar{T}=T$.

This last statement can be further analyzed studying the relation between biconnection $f(R)$ gravity and Brans-Dicke theories. It is well known that the action of both metric and Palatini $f(R)$ gravities can be related to the Brans-Dicke action employing conformal transformations \cite{Sotiriou:2008rp}. The same can be done with action (\ref{029}) when reduced to its $f(R)$ subcase.

Consider the following action
\begin{equation}
S=\int d^4x \sqrt{-g}\left[ f(\chi)+ \frac{\partial f(\chi)}{\partial \chi} \left(\Re-\chi\right)\right] \,,
\label{045}
\end{equation}
with $\chi$ a scalar field. Variation with respect to $\chi$ yields
\begin{equation}
\frac{\partial^2 f(\chi)}{\partial \chi^2} \left(\Re-\chi\right)=0 \,,
\end{equation}
which, provided $\partial^2 f/\partial \chi^2 \neq 0$, gives $\Re=\chi$.
This inserted into (\ref{045}) takes us back to the biconnection $f(R)$ action, showing that the two actions are physically equivalent, i.e.~they yield to the same equations of motion.
We can now define a new scalar field $\xi$ and its potential $V(\xi)$ as
\begin{equation}
\xi\equiv\frac{\partial f(\chi)}{\partial \chi} \quad\mbox{and}\quad V(\xi)\equiv f(\chi(\xi))-\xi\,\chi(\xi) \,,
\label{049}
\end{equation}
in terms of which action (\ref{045}) becomes
\begin{equation}
S=\int d^4x \sqrt{-g} \left[\xi\,\Re-V(\xi)\right] \,.
\end{equation}
Note that this action does not coincide with the Brans-Dicke one since we still have $\Re$ in place of $R$. However, thanks to (\ref{044}), we can expand $\Re$ and obtain
\begin{equation}
S=\int d^4x \sqrt{-g}\left\{\xi\,g^{\mu\nu} \left[R_{\mu\nu}(\Gamma) -\stackrel{\Gamma}{\nabla}_{[\mu} \mathcal{K}_{\nu\lambda]}{}^\lambda -S_{\mu\sigma}{}^\lambda(\Gamma) \mathcal{K}_{\nu\lambda}{}^\sigma \right] -V(\xi) \right\} \,.
\label{046}
\end{equation}

At this point we make use of condition (\ref{019}), and replace $\Gamma$ with $\tilde\Gamma$ in (\ref{046}). Since from (\ref{019}) we have $S_{\mu\nu}{}^\lambda(\tilde\Gamma)=0$, we obtain
\begin{equation}
S=\int d^4x \left[\sqrt{-g}\,\xi\,R(\tilde\Gamma) +\mathcal{K}_{\nu[\lambda}{}^\lambda \tilde{\nabla}_{\mu]}\left(\sqrt{-g}\,\xi\,g^{\mu\nu}\right) -\sqrt{-g}\,V(\xi)  \right] \,,
\label{047}
\end{equation}
where a surface term has been neglected.
Introducing the metric $h_{\mu\nu} \equiv F\,g_{\mu\nu}= \xi\,g_{\mu\nu}$, the connection $\tilde\Gamma$, as given by (\ref{019}), can be rewritten as
\begin{equation}
\tilde\Gamma_{\mu\nu}^\lambda = \frac{1}{2} h^{\lambda\sigma} \left(\partial_\mu h_{\nu\sigma}+\partial_\nu h_{\mu\sigma} -\partial_\sigma h_{\mu\nu}\right) \,.
\label{033}
\end{equation}
This shows that $\tilde\Gamma$ is the Levi-Civita connection with respect to $h_{\mu\nu}$, and in particular we have
\begin{equation}
\tilde{\nabla}_\mu\left(\sqrt{-h}\, h_{\alpha\beta}\right) =\tilde{\nabla}_\mu\left(\sqrt{-g}\,\xi\, g_{\alpha\beta}\right) =0 \,.
\end{equation}
Hence action (\ref{047}) reduces to
\begin{equation}
S=\int d^4x\sqrt{-g} \left[\xi\,R(\tilde\Gamma)-V(\xi)  \right] \,,
\end{equation}
which expanding $R(\tilde\Gamma)$ becomes
\begin{equation}
S=\int d^4x\sqrt{-g} \left[\xi\,R+\frac{3}{2\xi}\, \partial_\mu\xi\partial^\mu\xi -V(\xi)  \right] \,.
\label{048}
\end{equation}

Exactly as it happens with the Palatini approach, the biconnection $f(R)$ action is conformally equivalent to a Brans-Dicke theory with BD parameter $\omega_0=-3/2$.
This enforces the previous statement that in vacuum biconnection modified gravitional theories described by action (\ref{029}), reduce to their Palatini counterparts. However when matter is present in the theory, the field equations (\ref{021}) present additional terms which do not appear considering a Palatini variation. In the passages above, the matter action would change when we switch from $\Gamma$ to $\tilde\Gamma$ adding new source terms. In this respect, when matter is present, biconnection $f(R)$ gravity can be regarded as equivalent to a Brans-Dicke theory with a modified source.

\section{Conclusion}
\label{Sec:conc}

In the present work a new variational approach to GR and modified theories of gravity has been proposed. The independent degrees of freedom are represented by the metric tensor $g^{\mu\nu}$, and two independent connections $\Gamma_{\mu\nu}^\lambda$ and $\Omega_{\mu\nu}^\lambda$, on which no a priori conditions have been imposed. The well-known Einstein-Hilbert action has been generalized with the biconnection GR action $S_{BC}$, as given by (\ref{001}).
Setting $\Gamma=\Omega$ the Einstein-Hilbert action can be recovered from $S_{BC}$. 
It has been shown that variation of $S_{BC}$ with respect to the above independent degrees of freedom leads to the usual Einstein field equations.
Both the two independent connections do not appear in the final form of the field equations and thus can be regarded as auxiliary fields.

A matter action can be added to $S_{BC}$ where dependence upon one of the two independent connections can be kept completely general.
Irrespective of this dependence the field equations always reduce to the usual Einstein field equations.
In this respect the biconnection variational approach generalizes the Palatini variational principle where, though one always obtains the Einstein field equations, the constraints that the matter action has to be independend of the connection and that the connection itself has to be taken torsion free, have to be considered from the beginning. If these assumptions are relaxed the variational principle becomes the metric-affine one. However within this last variational approach the field equations do not naturally reduce to the usual Einstein field equations \cite{Hehl:1994ue}.

Then it has been considered a generalization of $S_{BC}$ in analogy with the common generalizations of the Einstein-Hilbert action given by $f(R)$ and Scalar-Tensor theories. Employing action (\ref{029}) these two classes of gravitational theories have been studied within a single analysis.
It emerged that the field equations of biconnection modified gravity result similar to their Palatini counterparts and reduce to these whenever the matter action do not depend on the connection. The most evident deviations depend on the (symmetric) hypermomentum $\sigma_{\mu\nu}{}^\lambda$ which vanishes for any known macroscopic matter field.

Exactly as the Palatini approach, the biconnection variational principle gives rise to different physical theories when applied to modified gravitational Lagrangians, while it leads to the usual Einstein field equations when the Einstein-Hilbert action is considered. In vacuum however the two variational approaches give rise to the same physical theory since the field equations coincide. This issue has been further studied under the point of view of equivalences with Brans-Dicke theories, where it has been shown that, employing conformal transformations, the action of biconnection $f(R)$ gravity is  equivalent to a Brans-Dicke theory with BD parameter $\omega_0=-3/2$, which is exactly the same theory equivalent to Palatini modified gravity.

\section*{Acknowledgments}

I would like to thank Tomi Koivisto for useful comments and suggestions about the paper.
I am deeply indebted to Christian B\"ohmer, not only for his willingness to help and to correct the paper, but also for on-going support and encouragement.

\appendix

\section{}
\label{app1}

In this appendix we derive the relation between $\bar{T}_{\mu\nu}$ and $T_{\mu\nu}$.

Consider the matter action
\begin{equation}
S_M(g,\Gamma,\Psi) = \int d^4x\, \mathcal L_M(g,\Gamma,\Psi) \,,
\end{equation}
with $\Psi$ denoting all the matter fields collectively (including $\phi$ if presents). Taking the variation with respect to $g^{\mu\nu}$ gives
\begin{eqnarray}
\delta_g S_M &=& \int d^4x \frac{\delta\mathcal L_M}{\delta g^{\mu\nu}} \delta g^{\mu\nu} \nonumber\\
&=&\int d^4x \left( \left.\frac{\delta\mathcal L_M}{\delta g^{\mu\nu}}\right|_\Gamma \delta g^{\mu\nu} +\frac{\delta\mathcal L_M}{\delta\Gamma_{\alpha\beta}^\lambda} \delta\Gamma_{\alpha\beta}^\lambda \right) \nonumber\\
&=&- \frac{1}{2} \int d^4x \sqrt{-g} \left[ \bar{T}_{\mu\nu}(\Gamma) \delta g^{\mu\nu} +2 \Delta_\lambda{}^{\alpha\beta} \delta\Gamma_{\alpha\beta}^\lambda\right] \,.
\label{032}
\end{eqnarray}
In GR we have
\begin{equation}
\delta_gS_M =\int d^4x \frac{\delta\mathcal L_M(g,\Psi)}{\delta g^{\mu\nu}} \delta g^{\mu\nu} \equiv -\frac{1}{2} \int d^4x \sqrt{-g} \,\delta g^{\mu\nu} T_{\mu\nu} \,,
\label{040}
\end{equation}
where in $\mathcal L_M(g,\Psi)$ the $\Gamma$s are substituted with Levi-Civita connections which are fully determined by the metric.
In order to find the relation between $T_{\mu\nu}$ and $\bar T_{\mu\nu}$, we have to replace $\Gamma$ with the Levi-Civita connection in (\ref{032}). 
The variation of the Levi-Civita connection with respect to the metric is given by \cite{Pal19}
\begin{equation}
\delta\Gamma_{\alpha\beta}^\lambda = \delta\left\{_{\alpha\beta}^{\,\,\lambda}\right\} = \frac{1}{2}g^{\lambda\sigma} \left(\nabla_\alpha\delta g_{\beta\sigma} +\nabla_\beta\delta g_{\alpha\sigma} -\nabla_\sigma\delta g_{\alpha\beta}\right) \,.
\end{equation}
Inserting this in (\ref{032}), considering that $\delta\Gamma_{\alpha\beta}^\lambda$ is now symmetric in the lower indices, and integrating by parts yields
\begin{eqnarray}
\delta_gS_M &=& -\frac{1}{2}\int d^4x\sqrt{-g}\bigg[\bar{T}_{\mu\nu}\delta g_{\mu\nu} -\delta g_{\beta\sigma}\nabla_\alpha\sigma^{\sigma\alpha\beta}  -\delta g_{\alpha\sigma}\nabla_\beta\sigma^{\sigma\alpha\beta} +\delta g_{\alpha\beta}\nabla_\sigma\sigma^{\sigma\alpha\beta}\bigg] \nonumber\\
&=&\frac{1}{2}\int d^4x\sqrt{-g}\, \delta g_{\mu\nu} \left[\bar{T}{}^{\mu\nu} +\nabla_\lambda\left(\sigma^{\nu\lambda\mu}+\sigma^{\nu\mu\lambda} -\sigma^{\lambda\mu\nu}\right) \right] \nonumber\\
&=&-\frac{1}{2}\int d^4x\sqrt{-g}\, \delta g^{\mu\nu} \left[\bar{T}_{\mu\nu} +\nabla^\lambda\left(\sigma_{\nu\lambda\mu}+\sigma_{\nu\mu\lambda} -\sigma_{\lambda\mu\nu}\right) \right] \,,
\end{eqnarray}
where surface terms have been discarded.
Finally, comparing this result with the definition of the GR energy-momentum tensor given in (\ref{040}), yields
\begin{equation}
T_{\mu\nu} =\bar{T}_{\mu\nu} +\nabla^\lambda\left(\sigma_{\nu\lambda\mu}+\sigma_{\nu\mu\lambda} -\sigma_{\lambda\mu\nu}\right) \,.
\label{041}
\end{equation}
So the term appearing under the covariant derivative in (\ref{041}) represents the part coming from the variation of the connection with respect to the metric.

\end{document}